# 基于分形优化变分模态分解的电力系统谐波间谐波检测

裴宇航[1]，喻敏[1,2]，余艳[1]

(1. 武汉科技大学理学院　湖北　武汉　430065;
2. 冶金工业过程系统科学湖北省重点实验室(武汉科技大学)　湖北　武汉　430065)

**摘要**：针对变分模态分解(Variational mode decomposition, VMD)的分解层数需事先人为确定的问题,首次提出了基于VMD分解分量分形盒维数(Fractal box dimension, FBD)最小值的参数确定方法。该方法首先对原始电力信号进行变分模态分解,选择VMD分解的分量中分形盒维数最小的K值作为VMD分解的层数,获得既定的若干内蕴模态函数(Intrinsic Mode Functions, IMFs),得到的IMFs即为电力系统的基波、谐波、间谐波信号。再进行希尔伯特变换(Hilbert Transform, HT),提取谐波信号的瞬时幅值、瞬时频率等特征。将该方法应用于电力系统仿真信号与实际数据,分析分解分量与真实分量的相对误差,并与经验模态分解(Empirical Mode Decomposition, EMD)、集合经验模态分解(Ensemble Empirical Mode Decomposition, EEMD)方法进行对比,实验结果表明该方法较于能更准确识别谐波间谐波信号,并实现有效提取。

**关键词**：变分模态分解;分形盒维数;谐波间谐波;希尔伯特变换

## Harmonic and Interharmonic Detection in Power Systems Based on Fractal-Optimized Variational Mode Decomposition

PEI Yuhang[1], YU Min[1,2], YU Yan[1]

(1.College of Science Wuhan University of Science and Technology Wuhan 430065 China;
2. Hubei Province Key Laboratory of Systems Science in Metallurgical Process,Wuhan University of science and Technology,Wuhan,430065 China )

**Abstract:** The proposed method introduces a parameter determination approach based on the minimum Fractal box dimension (FBD) of Variational Mode Decomposition (VMD) components, aiming to address the issue of manual determination of VMD decomposition layers in advance. Initially, VMD is applied to the original power signal, and the layer number for VMD decomposition is determined by selecting the K value associated with the smallest fractal box dimension among its components. Subsequently, several Intrinsic Mode Functions (IMFs) are obtained as fundamental, harmonic, and interharmonic signals representing different aspects of the power system. Furthermore, Hilbert transform(HT) is employed to extract instantaneous amplitude and frequency information from these harmonic signals. Experimental evaluation using simulation data and real-world power system data demonstrates that compared to Empirical Mode Decomposition (EMD) and Ensemble Empirical Mode Decomposition (EEMD), our proposed method achieves more accurate identification and effective extraction of harmonic signals.

This work is supported by National Natural Science Foundation of China (No. 61671338).

## 0 引言

电力系统中的谐波是一种不可避免的现象,它源于非线性负载、电力电子设备以及其他系统组件引起的波形失真。随着电力系统的不断发展和复杂化,对谐波的准确检测变得尤为重要[1-4]。

傅里叶变换[5-7]、小波变换[8-14]、希尔伯特-黄变换(Hilbert-Huang Transform, HHT)[15-19]等是常用的谐波间谐波时频分析方法。傅里叶变换具有较快的检测速度,但只适用于确定性的平稳信号,不易检测时变的谐波信号[20]。此外,由于傅里叶变换难以确定谐波波形周期,易于受非同步采样影响而产生"栅栏效应"和"频谱泄漏"等问题[21]。小波变换在时频两域都有表征信号局部特征的能力,但需要确定小波基和分解层数,不同的小波基和分解层数检测同一谐波信号得到的结果差别可能会很大,且小波分解进行谐波分解时存在频带混叠现象。HHT是近年来分析非线性波形时性能较好的时频工具,其分别由经验模态分解(EMD)与希尔伯特变换(Hilbert Transform, HT)两部分构成。HHT通过将信号分解成固有模态函数(IMFs)和希尔伯特谱进行分析,能够较好地处理非线性和非平稳信号。因其无需选择基函数且能进行自适应分解诶,近年来被广泛应用于电力系统谐波检测中。然而,EMD分解方法易产生虚假分量,并且会产生模态混叠和端点效应等问题,同时也无法解释HT产生的负频率。为了克服这些问题,由Zhaohua Wu和Norden E. Huang于2009年提出了集合经验模态分解(EEMD)[22]。EEMD具有能量保持性好、适应性强等特点,有效解决了传统经验模态分解的模态混叠问题,最大程度保留真实信号。但同时由于其计算复杂度较高,尤其在处理长时间序列时,会消耗大量计算资源。

变分模态分解[23](Variational Mode Decomposition, VMD)是信号处理中一种较新的时频分析方法,其具有坚实的数学理论基础,且分解精度较高、计算速度较快,因此被应用于诸多领域。但实验证明采用VMD进行谐波检测时,若分解层数K值选取不当,VMD易出现过分解、模态混叠现象,从而影响谐波检测结果。文献[24]运用频谱分析的方法来确定VMD分解的模态数,但需人为观察频谱个数。因此为了通过VMD准确提取谐波间谐波信号,本文提出了一种基于分形优化VMD算法中参数K的谐波检测方法,将每个K值下对应VMD分解的各模态的分形盒维数作为其分形盒维数值。当K等于真实模态数目时,其对应的分形盒维数值就会达到最低点且趋于平稳,因此选择分形盒维数值最小的K值作为VMD分解层数K。仿真信号和真实信号实验均表明:该方法较好地解决了需人为设定VMD中参数K的问题,参数优化效果好。

# 1 基础理论

## 1.1 变分模态分解(VMD)

VMD是由Dragomiretshiy于2013年提出的一种全新的自适应、非递归信号分解方法[25,26]。该方法不仅解决了EMD所存在的模态混淆和端点效应问题,同时有效地减少了伪分量和模态混淆的现象。相比之下,它也弥补了小波分解无法自适应地处理信号的不足。在经过计盒维数算法优化的条件下,VMD的模态分量数量较少,分解效果更佳,且对噪声具有更强的鲁棒性[27]。VMD算法主要分为两部分,即变分问题的构造与变分问题的求解。

### 1.1.1 构造变分问题

构造原理为:搜索使得每个模态函数的估计带宽之和最小的$K$个模态函数$u_k(t)$,VMD通过以下三个步骤分解每一个模态函数。

**步骤1** 采用Hibert变换计算每个模态函数$u_k(t)$的解析信号以获得其单边频谱。

$$(\delta(t)+\frac{j}{\pi t})*u_k(t) \quad (1)$$

**步骤2** 通过各模态解析信号对应中心频率$e^{-j\omega_k t}$调整各模态使其频谱调制到基频带上。

$$[(\delta(t)+\frac{j}{\pi t})*u_k(t)]e^{-j\omega_k t} \quad (2)$$

**步骤3** 根据高斯平滑度和梯度平方准则解调信号,并计算信号梯度的平方$L^2$范数作为该模态函数带宽的估计。分解后的各模态量均为调幅-调频信号,变分约束问题如下:

$$\min_{u_k,\omega_k}\left\{\sum_k\|\partial_t[(\delta(t)+\frac{j}{\pi t})*u_k(t)]e^{-j\omega_k t}\|_2^2\right\} \quad (3)$$

$$s.t.\sum_k u_k=f(t)$$

### 1.1.2 求解变分问题

为求解上述变分约束问题最优解,引入二次惩罚项和拉格朗日乘数$\lambda$,惩罚因子$\alpha$,二次惩罚因子通过限定模态带宽保证信号重构精度,从而使得变分约束问题转化为无约束的优化问题:

$$\begin{aligned}L(u_k,\omega_k,\lambda)=&\alpha\sum_k\|\partial_t[(\delta(t)+\frac{j}{\pi t})*u_k(t)]e^{-j\omega_k t}\|_2^2\\&+\|f(t)-\sum_k u_k(t)\|_2^2\\&+\langle\lambda(t),f(t)-\sum_k u_k(t)\rangle\end{aligned} \quad (4)$$

其中$f(t)$为原始信号,VMD采用交替方向乘子法(ADMM)求解式(1),并通过交替更新

$u_k^{n+1}, \omega_k^{n+1}, \lambda_k^{n+1}$ 寻找每个模态分量的中心频率和带宽从而找到约束变分模型的最优解。

求解 $u_k^{n+1}, \omega_k^{n+1}$ 的优化方程为：

$$u_k^{n+1} = \arg\min_{u_k \in X} \{\alpha \| \partial_t[(\delta(t)+\frac{j}{\pi t})*u_k(t)]e^{-j\omega_k t} \|_2^2 \\ + \| f(t)-\sum_i u_i(t)+\frac{\lambda(t)}{2} \|_2^2\} \quad (5)$$

$$\omega_k^{n+1} = \arg\min_{u_k \in X} \{\alpha \| \partial_t[(\delta(t)+\frac{j}{\pi t})*u_k(t)]e^{-j\omega_k t} \|_2^2 \quad (6)$$

得到的最优解下模态分量新 $u_k, \omega_k$ 分别为：

$$\hat{u}_k^{n+1}(\omega) = \frac{\hat{f}(\omega)-\sum_{i\neq k}\hat{u}(\omega)+\frac{\hat{\lambda}(\omega)}{2}}{1+2\alpha(\omega-\omega_k)^2} \quad (7)$$

$$\omega_k^{n+1} = \frac{\int_0^\infty \omega|\hat{u}_k(\omega)|^2 d\omega}{\int_0^\infty |\hat{u}_k(\omega)|^2 d\omega} \quad (8)$$

## 1.2 分形盒维数(FBD)

分形维数是分形学中评估分形对象复杂度、不规则性和自相关性的重要度量标准,用于确定其非整数维空间维度。此外,分形维数还常用于描述混沌吸引子的内在相似性。因此,通过分形维数可以反映模态分量的自相关性,进而评估是否存在模态混叠现象。分形维数值越高,模态自相关性越强,模态混叠越严重。因此,当信号分解效果良好时,各模态分量的分形维数将逐渐减小并逐步趋于稳定。

在经典的分形维数中,包括盒维数、信息维数和关联维数等。不同的计算方法会导致模态分量的分形维数值有所不同。由于计算盒维数的方法简单易行,并为其他维数概念提供了衍变的基础,因此本文选择计算盒维数来研究风速信号中各模态分量的分形维数值。

其原理是用边长为ε的正方体盒子覆盖观测物,设观测物面积为 $S$，$N(\varepsilon)$ 为所需的最少盒子个数,则：

$$N(\varepsilon) = S/\varepsilon^2 \quad (9)$$

两边取对数得：

$$\ln N(\varepsilon) = \ln S + 2\ln(1/\varepsilon) \quad (10)$$

则计盒维数 $D(Z)$ 记为：

$$D(Z) = \lim_{\varepsilon\to 0}\frac{\ln N(\varepsilon)}{\ln(1/\varepsilon)} \quad (11)$$

当ε趋于 0 时,$N(\varepsilon)$ 随着ε减小而增加的对数速率即为计盒维数。计算不同 K 值下 VMD 分解的模态分量盒维数,取模态分量盒维数最早趋于稳定时所对应的 K 值为 VMD 分解最优分解层数。

## 2 基于分形优化变分模态分解的谐波检测

### 2.1 不同 K 值对 VMD 分解性能的影响

VMD 对非平稳、非线性信号具有良好的分解效果。但同许多经典聚类和分段算法(如 K-means)一样,其分解效果受参数选取的影响较大。如果分解参数 K 设置不合理,则会导致分解结果出现较大误差。

由 VMD 的约束条件可知,VMD 分解的本质是使得输入信号近似等于 K 个模态之和,因而当 K 过小时,易导致信号分解不足,信号中的部分分量混在一起,呈现欠分解状态。而 K 过大则会使得信号分解过度,信号的重要部分由两个或多个不同的模式共享,并且它们的中心频率重合,呈现过分解状态。因此在使用 VMD 进行信号分解时,需预先确定合适的 K 值。

为更直观地说明,给定信号如式(12)所示,K 取不同值时,对该信号进行 VMD 分解,分解结果如图 1 所示。

$$f(t) = \begin{cases} \cos(6\pi t), & 0 \leq t < 1 \\ \frac{1}{4}\sin(56\pi t), & 0 \leq t < 1 \\ \frac{1}{16}\cos(542\pi t), & 0 \leq t < 1 \end{cases} \quad (12)$$

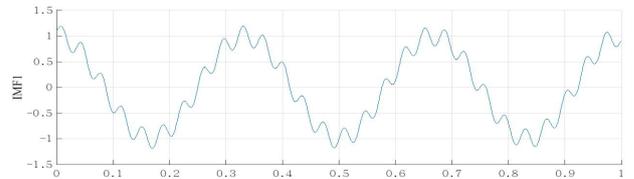
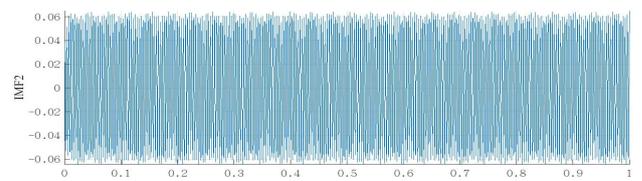

K=2

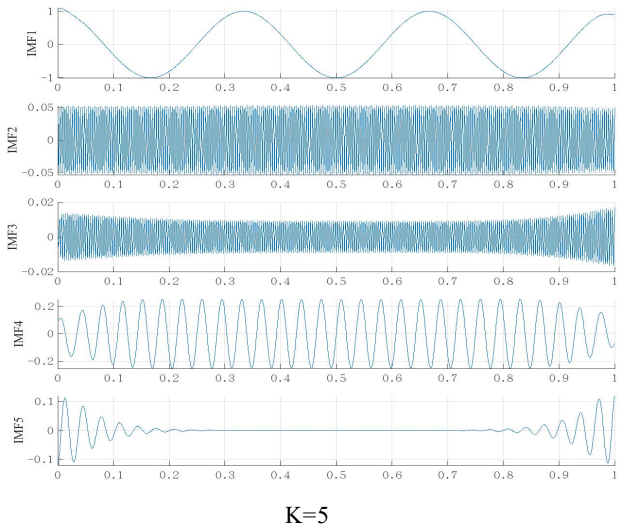

K=5

图 1 不同 K 值下信号 VMD 分解图

Fig.1 VMD decomposition diagram of signals under different K values

由图 1 可知,K=2 时,IMF1 中明显混有多个频率分量,导致波形畸变,说明 K 过小,信号未被完全分解;K=5 时,IMF2、IMF4 频率相同,说明 K 过大,导致一个信号成分被分解为多个,信号被过度分解,而且出现虚假分量 IMF5。由此可知 K 值不同,得到的结果差异很大。

### 2.2 基于分形值优化参数 K

在电力系统中,当出现谐波信号和间谐波信号时,针对很难确定 K 值从而提取谐波和间谐波信号特征的问题,提出了基于分形值优化 VMD 参数 K 的信号分解方法。

当产生谐波信号时,整个电力信号的复杂度会因为混杂了谐波信号和间谐波信号而增加,通过分形盒维数的值进行衡量信号的复杂度是一个较好的方法。进行 VMD 分解时,我们从 K=1 开始,逐渐使 K 值增大,取当前分解模态中分形盒维数的最小值代表当前 K 所对应的分形盒维数值。当分解层数 K 小于真实的模态数目时,会出现模态混叠的现象。因此在当前 K 值下,分形盒维数的值就会很高。当 K 值逐渐增大时,模态混叠的情况会越来越少。直到 K 值等于真实模态数目时,K 值对应的分形盒维数的值就会达到最低点,并且再当 K 值增大时,K 值对应的分形盒维数值会变得平稳。因此当分解层数 K 等于真实的模态数时,就会产生一个"拐点",这个"拐点"就是我们优化后的 K 值。

### 2.3 算法步骤以及流程

本文提出的基于分形盒维数优化 VMD 的谐波检测方法,旨在利用 VMD 优于 EMD 的性能,从原始信号的 VMD 分解,得到相应的 IMF 分量,从而检测出原始信号中的谐波和间谐波分量。其检测流程图,如图 2 所示。

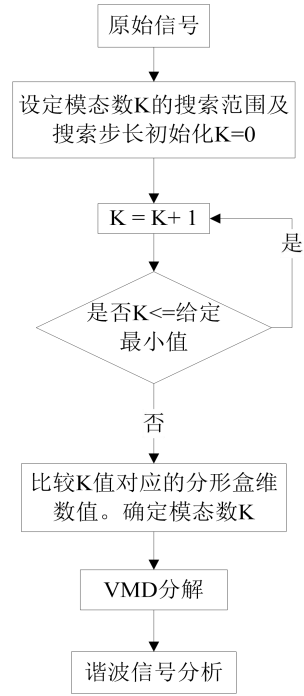

图 2 检测方法流程图

Fig. 2 Flow chart of detection method

具体步骤如下:

**步骤 1** 首先,将模态数 K 初始化为 1,并采用默认参数设置:惩罚因子 α = 4096、带宽 τ = 0、直流成分 DC = 0、初始参数 init = 1,并设置容差 tol = $10^{-10}$。

**步骤 2** 对电力系统信号进行 VMD 分解。在这一过程中,可能会出现类似残差的信息量较少的虚假分量。这些虚假分量的分形盒维数值较小,可能会对算法的准确性产生影响。因此,在计算各个模态的分形盒维数值之前,首先移除这些类虚假残差分量。然后,计算每个模态下的分形盒维数值,并选择当前 K 值下每个模态中的最小分形盒维数值作为该 K 值的分形盒维数值。

**步骤 3** 逐步增加 K 值,直至达到预设值。在每一步中,都要重复执行步骤 2 以获取当前 K 值相应的分形盒维数值。

**步骤 4** 比较不同 K 值下的分形盒维数值。

选择具有较小分形盒维数值且趋于稳定的"拐点"所对应的 $K$ 值作为最终的 VMD 分解的 $K$ 值。通过这一步骤获得的 IMF 分量可用于进行谐波检测和分析,从而揭示电力系统中的谐波特性。

## 3 仿真与实际数据分析

**算例1:稳态谐波间谐波检测**

电力系统中含间谐波的稳态仿真信号模型可表示为式(13):

$$f(t) = \sum_{m=1}^{M} A_m \sin(2\pi f_m t + \varphi_m) \quad (13)$$

式中,$A_m$、$f_m$ 和 $\phi_m$ 分别为各谐波分量的幅值、频率和相位信息。

本文采用了 IEEE 间谐波工作组里的例子[28]对本文所提方法进行验证分析,其包含 1 个基波、四个间谐波、1 个 5 次谐波,如式(14)所示。采样频率为 4096Hz,采样点为 4096。

$$f(t) = \begin{cases} \sin(100\pi t), & 0 \leq t < 1 \\ 0.3\sin(208\pi t), & 0 \leq t < 1 \\ 0.4\sin(234\pi t), & 0 \leq t < 1 \\ 0.2\sin(268\pi t), & 0 \leq t < 1 \\ 0.2\sin(294\pi t), & 0 \leq t < 1 \\ 0.5\sin(500\pi t), & 0 \leq t < 1 \end{cases} \quad (14)$$

对信号进行 VMD 分解,初始化 $K = 1$,设定 $K$ 的搜索范围为[1, 10],利用分形盒维数最小值原则优化模态 $K$,计算各个 $K$ 值对应的分形盒维数值。模态数与分形盒维数值的关系图,如图 3 所示。由图可知,当 $K = 7$ 时,$K$ 值对应分形盒维数值开始趋于稳定,其值为 1.0193。所以取 $K = 7$ 作为 VMD 分解层数 $K$ 的最优值。

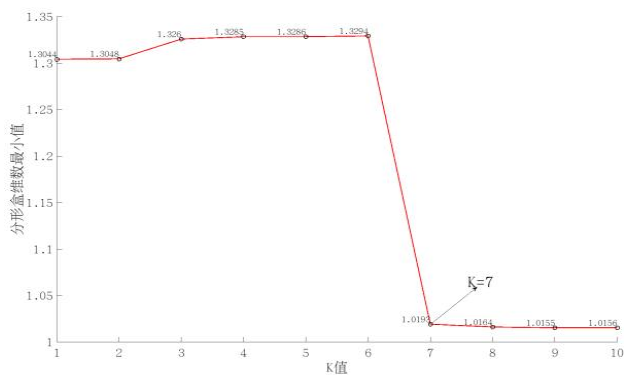

图 3 原始信号 K 值对应分形盒维数值

Fig.3 The original signal K value is matched to the fractal box dimension

VMD 分解 $K$=7 的结果、EMD、EEMD 分解结果如图 4-6 所示。

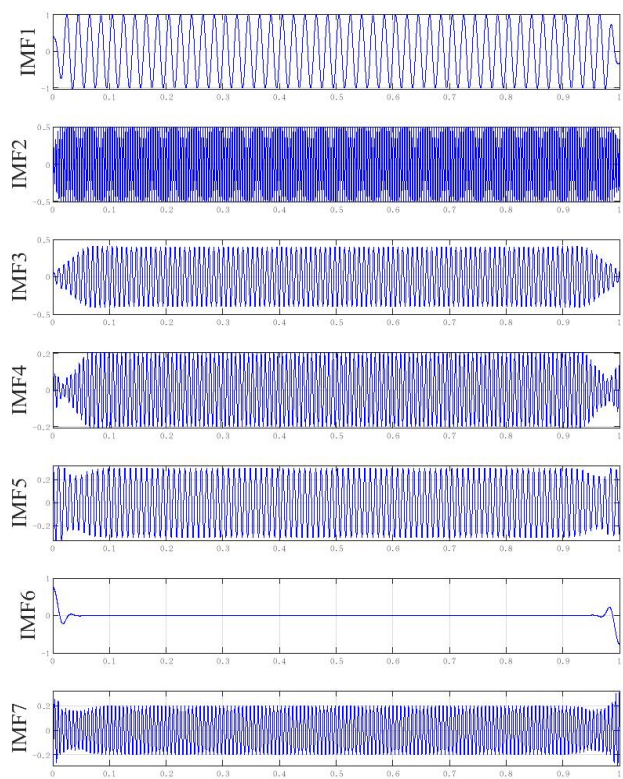

图 4 原始信号 VMD 分解

Fig.4 VMD decomposition of the original signal

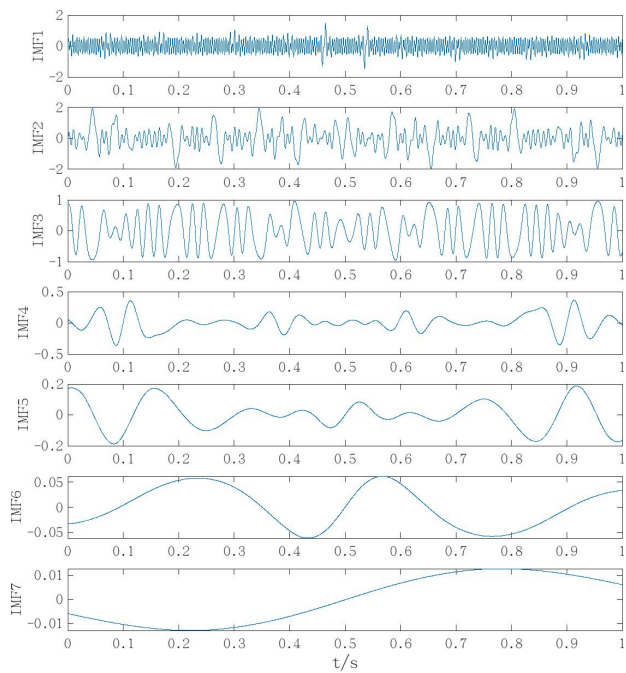

图 5 原始信号 EMD 分解

Fig.5 EMD decomposition of the original signal

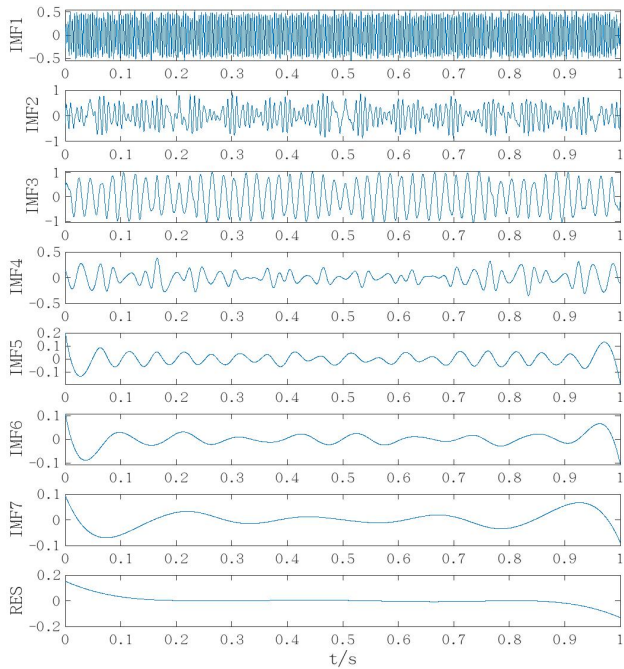

图 6 原始信号 EEMD 分解
Fig.6 EEMD decomposition of the original signal

由图 4 可以看出 VMD 分解得到的 IMF1-IMF5、IMF7 均呈现为正弦波动信号,没有发生畸变,这说明这些分量为单一的谐波间谐波分量,并且原始信号中的 6 个分量均能与其一一对应。这说明进行参数优化后的 VMD 分解方法能准确的分解出原始信号的各个分量。从图 5 对应的 EMD 分解结果中可以看到,EMD 分解方法连基波信号都没能分解出来,仅能看出 IMF1 为 5 次谐波分量,其余间谐波信号也均未能分解出来。而图 6 中 EEMD 方法仅能分解出 IMF3 为基波、IMF1 为 5 次谐波,且分解出来的基波幅值有一定的降低,其余间谐波分量亦未能分解。不仅如此,IMF2 出现了明显的模态混叠现象,IMF5 为较难判断的虚假分量。究其原因,是由于原始信号中的间谐波频率较近,而 EMD 与 EEMD 方法仅能分解出较大频率与较小频率比值大于 2[29]的信号。显然当前算例中的间谐波信号并未满足上述条件,因此难以得到这些间谐波分量。同时,从图 5、6 可以看出 EMD 和 EEMD 产生了许多虚假分量和前分解分量,会影响判断。VMD 分解仅产生了一个虚假分量,并且产生的虚假分量基本为零,易于判断。

对上述信号进行 VMD 分解得出来的 6 个模态分量进行 Hilbert 变换得到各个分量的瞬时幅值与瞬时频率,并对其求得平均值,所得检测结果如表 1 所示。

表 1 稳态谐波间谐波信号检测结果
Tab.1 Test results of steady-state harmonic and interharmonic signals

| 幅值/V | | | 频率/HZ | | |
| --- | --- | --- | --- | --- | --- |
| 真值 | 检测值 | 误差比 | 真值 | 检测值 | 误差比 |
| 1.0 | 0.9843 | 0.0157 | 50 | 49.8947 | 0.0021 |
| 0.5 | 0.4966 | 0.0068 | 250 | 250.8001 | 0.0032 |
| 0.4 | 0.3805 | 0.0488 | 117 | 117.7771 | 0.0066 |
| 0.2 | 0.1979 | 0.0105 | 147 | 147.8232 | 0.0056 |
| 0.3 | 0.2948 | 0.0173 | 104 | 103.8625 | 0.0013 |
| 0.2 | 0.1893 | 0.0535 | 134 | 134.8086 | 0.0060 |

由表 1 可知,检测得到的各分量的幅值与频率均较精确,检测误差小,接近真实值。

以上结果表明:当电力信号中存在谐波和有较为接近的间谐波分量时,本文所提出的 VMD 参数优化 K 值方法适用且有效,优化参数后的 VMD 分解方法能将各分量信息都准确提取出来,且具有很高的检测精度。

**算例 2:时变有噪声谐波间谐波检测**

暂态谐波检测涉及到电力系统中出现的短时、突发性谐波事件。这可能由于设备的开关操作、突发性负载变化或其他不稳定因素引起。在这种情况下,谐波信号往往混杂在系统的动态变化中,而且噪声的影响可能更加显著。本文采用了下列暂态有噪声谐波信号对本文所提算法进行验证,其参数信息如式(15)所示。

$$f(t)=\begin{cases} sin(30\pi t), & 0 \leq t < 1.0 \\ 4sin(100\pi t), & 0 \leq t < 1.0 \\ 2sin(238\pi t), & 0.2 \leq t < 0.5 \\ 3sin(500\pi t), & 0.6 \leq t < 1.0 \end{cases} \quad (15)$$

在上述信号中加入了加入 38dB 的高斯噪声,用本文提出的方法对信号进行 VMD 分解,初始化 K=1,设定 K 的搜索范围为[1, 8],利用分形盒维数最小值原则优化模态 K。模态数与分形盒维数值的关系图如图(7)所示。由图可知,当 K=7 时,分形盒维数最小值趋于稳定并且为最低点,最小值为 1.1614。所以取 K=7 作为 VMD 分解的模态数。

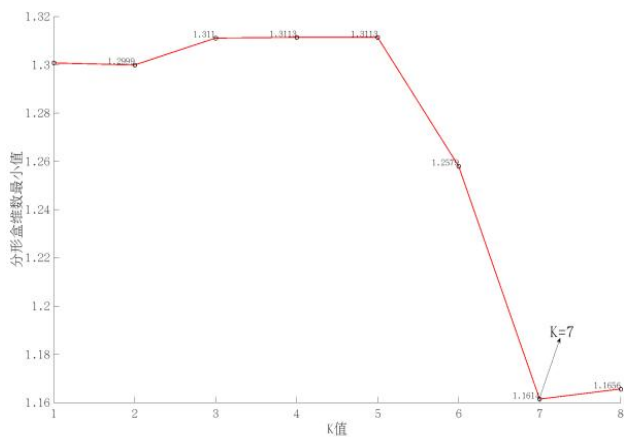

图 7 时变信号 K 值对应分形盒维数值

Fig.7 The K value of the time-varying signal corresponds to the fractal box dimension

VMD 分解 $K=7$ 的结果、EMD、EEMD 分解结果如图 8-10 所示。

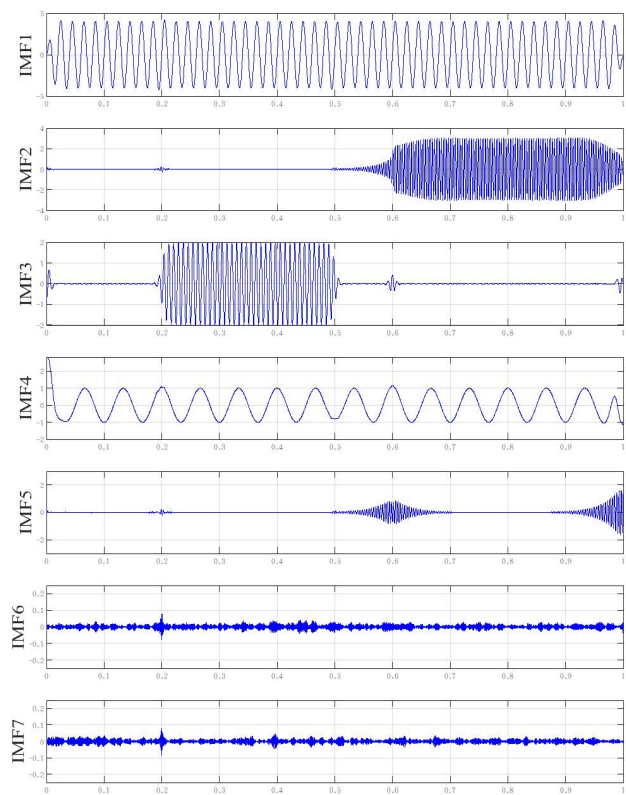

图 8 时变信号 VMD 分解

Fig.8 VMD decomposition of time-varying signals

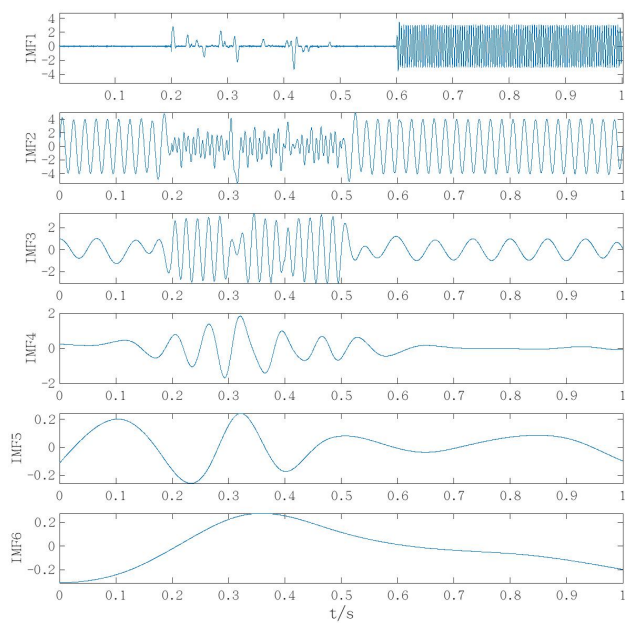

图 9 时变信号 EMD 分解

Fig.9 EMD decomposition of time-varying signals

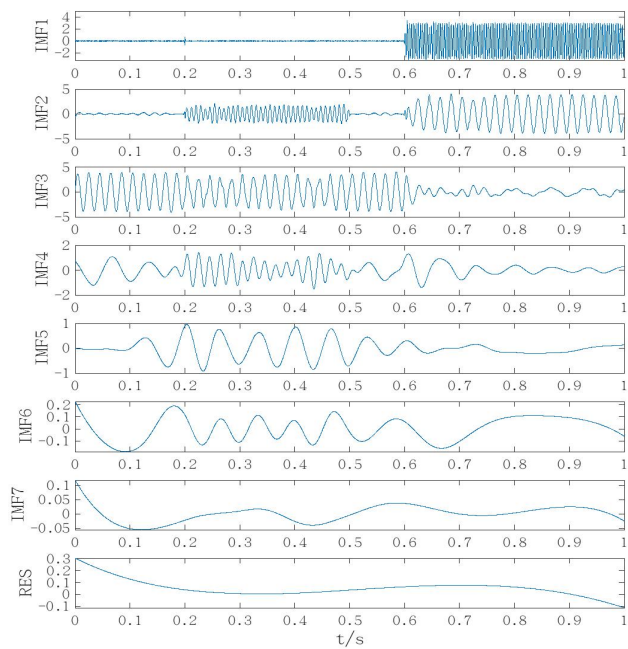

图 10 时变信号 EEMD 分解

Fig.10 EEMD decomposition of time-varying signals

由图 8 可以看出优化后的 VMD 分解中前四项可以对应原始时变信号中的各个分量。其中 IMF1 对应基波,IMF2 对应 0.6~1.0 时段的 5 次谐波,IMF3 对应 0.2~0.5 时段的 119Hz 间谐波,IMF4 对应 0~1.0 时段的 15Hz 间谐波。由此可见,优化后的 VMD 方法成功地将每一时段的信号分量均一

一分解。在图 9 中,IMF1 对应 5 次谐波;IMF2 为基波,在 0.2~0.5 时段为 119Hz 间谐波,但是可以看出存在模态混叠;IMF3 的 0.2~0.5 时段为基波,其他时段为 15Hz 间谐波,说明也发生了模态混叠现象。因此,EMD 方法未能较好的检测谐波间谐波信号。在图 10 中,IMF1 对应 5 次谐波;IMF2 按时段先后分别对应 119Hz 间谐波和基波,出现了模态混叠;IMF3 前 0~0.6 时段对应基波;又可以看出 IMF5 与 IMF6 的频率大致相同,为 15Hz 间谐波,说明存在过分解现象。因此,EEMD 方法同样未能较好的检测谐波间谐波信号。

对上述信号进行 VMD 分解得出来的前 4 个模态分量进行 hilbert 变换得到各个分量的瞬时幅值与瞬时频率,并对其求得平均值,所得检测结果如表 2 所示。

表 2 暂态时变有噪声谐波间谐波信号检测结果
Tab.2　Test results of transient time-varying noise between harmonic signals

| 幅值/V | | | 频率/HZ | | |
|---|---|---|---|---|---|
| 真值 | 检测值 | 误差比 | 真值 | 检测值 | 误差比 |
| 4 | 3.9521 | 0.0120 | 50 | 49.9695 | 0.0006 |
| 3 | 2.8277 | 0.0574 | 250 | 249.6672 | 0.0013 |
| 2 | 1.9676 | 0.0162 | 119 | 119.6031 | 0.0051 |
| 1 | 1.0241 | 0.0241 | 15 | 15.9012 | 0.0601 |

由表 1 可知,检测得到的各分量的幅值与频率均较精确,检测误差小,接近真实值。

以上结果表明:当电力信号中出现时变暂态间谐波且存在噪声时,本文所提出的 VMD 参数优化 K 值方法适用且有效,优化参数后的 VMD 分解方法能将各分量信息都准确提取出来。

**算例 3:实际谐波信号分析**

将基于分形优化的 $K$ 值的变分模态分解方法应用于实际信号 $v(t)$ 的成分提取中,该信号源于 220KV 变电站电能质量监测仪所获主变高压测的 C-相电压实测数据[30],采样频率等于 10240 HZ,采样点数等于 2048,为 10 个周波,即 0.2s。$v(t)$ 图 11 所示:

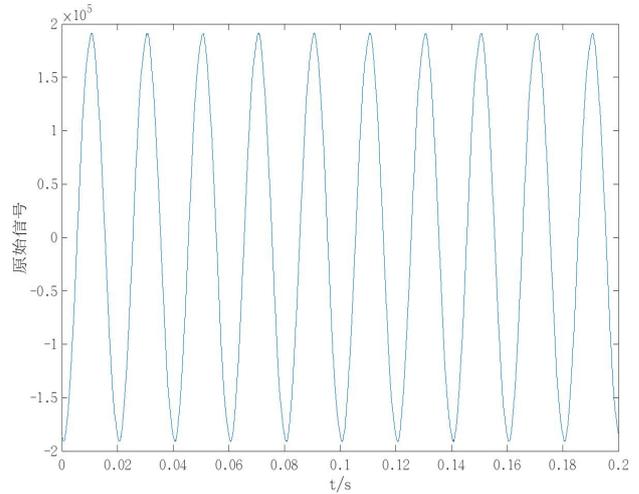

图 11　实际信号
Fig.11　Real signal

将本文提出的方法用于该实际信号,选取惩罚因子 $a = 10240$,$K$ 值搜索范围为[1,10],计算出的各个 $K$ 值对应的分形维数值如图 12 所示:

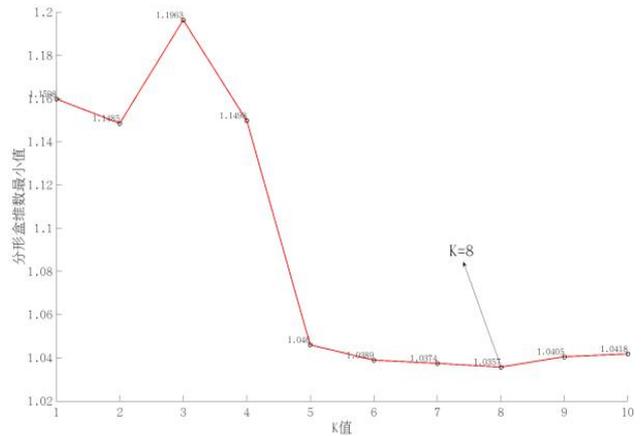

图 12　实际信号 K 值对应分形盒维数值
Fig.12 The K value of the real signal corresponds to the fractal box dimension value

因此可以看出,对实际信号进行 VMD 分解的 $K$ 值取为 8,其 VMD 分解结果、EMD、EEMD 分解结果如图 13-15 所示。

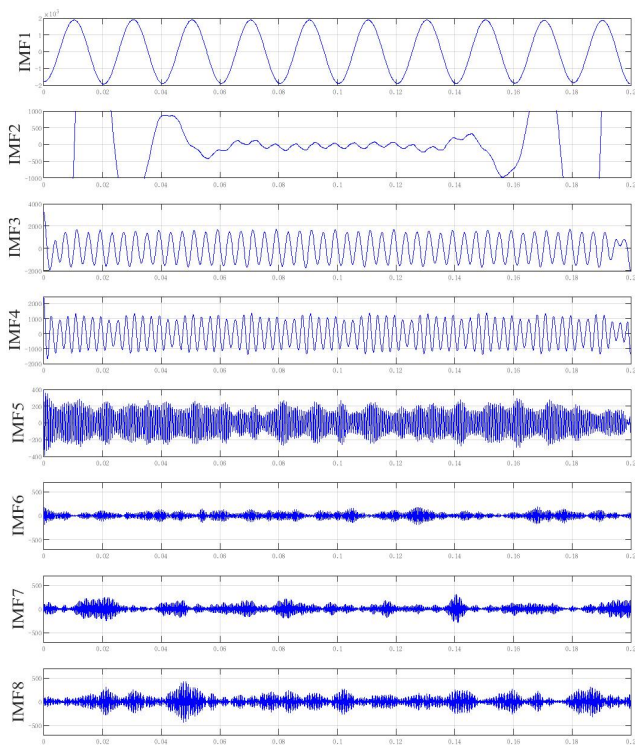

图 13 实际信号 VMD 分解
Fig.13 VMD decomposition of the real signal

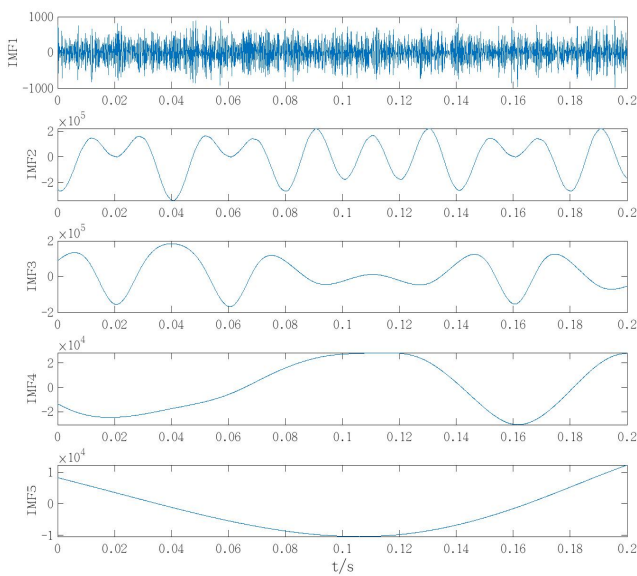

图 14 实际信号 EMD 分解
Fig.14 EMD decomposition of the real signal

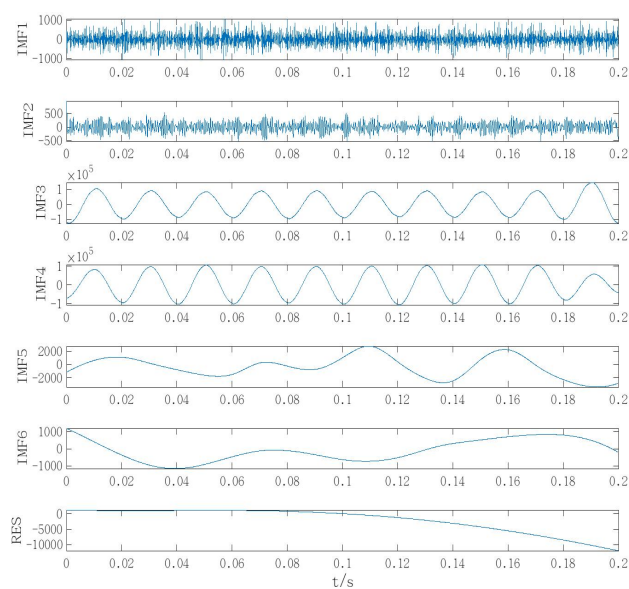

图 15 实际信号 EEMD 分解
Fig.15 EEMD decomposition of the real signal

从图 13 可以看出,原始信号经过参数优化后的 VMD 分解方法分解之后,得到的 IMF1 即为频率为 50HZ 的基波。同时文献[30]指出:该真实信号存在丰富的谐波,主要为 5 次谐波。经过 Hilbert 变换,得到 IMF3 的平均频率为 249.3HZ,即为五次谐波。同时经过 Hilbert 变换,得到 IMF4、IMF5 的平均频率为 354.2HZ、1549.2HZ,依次为 7 次谐波、31 次谐波。由波形图可以看出 IMF2 为一个暂态的谐波,在 0.06s 到 0.14s 内进行 Hilbert 变换得到 IMF2 平均频率为 105.4HZ,因此是一个暂态的间谐波。在图 14 中,该信号经过 EMD 分解后,IMF1 呈现出很强的模态混叠现象,IMF2 模态混叠也较严重,并且没有分解出频率为 50HZ 的基波,分解效果很差,无法得到有效的信息。在图 15 中,经过 EEMD 分解之后,IMF1 和 IMF2 同样呈现出严重的模态混叠现象,并且 IMF3 和 IMF4 频率相同大致相同,说明出现了过分解现象。这说明 EMD 和 EEMD 无法针对真实的谐波信号进行检测。

从以上实验结果说明本文提出的方法对真实的电力系统的谐波间谐波信号检测适用且有效,同时具有很高的分解精度,对比 EMD、EEMD 方法具有优越性。

## 4 结论

本文针对 VMD 方法中需要手动设定的模态

参数 $K$ 的问题,提出了一种新的方法,即基于分形优化 $K$ 值的变分模态分解方法。通过利用 VMD 分解信号的能力,该方法可以有效地将电力信号中的基波和谐波分解出来,并结合 Hilbert 变换来提取谐波的幅值和频率特征,从而实现对电力系统中稳态和暂态谐波信号的检测。通过分析与实验验证,得出了以下结论。

通过计算不同 $K$ 值下各模态的分形盒维数,并以最小值作为 $K$ 值的分形盒维数值,选取分形盒维数最小值对应的 $K$ 值优化 VMD 参数 $K$,成功解决了传统变分模态分解算法中需要手动设定 $K$ 值的问题,从而实现了信号的最优分解。

基于分形优化参数的变分模态分解能够有效地检测出电力信号中频率接近的谐波成分,适用于稳态和暂态谐波信号的检测。该方法具有很强的抗噪性,并且具有较高的检测精度,优于传统的 EMD 和 EEMD 分解方法。

经过参数优化后,VMD 分解得到的虚假分量数量明显少于 EMD 和 EEMD 分解,并且这些虚假分量的幅值较小,能够与真实分量区分开来,因此可以大大降低误判的可能性。

**参考文献**

作者简介：
　　裴宇航(2003-),男,本科生,研究方向为电力系统谐波间谐波检测等。E-mail:peiyuhang2@gmail.com
　　喻　敏(1975-),女,通信作者,博士研究生,讲师,研究方向为分形与小波应用、电能质量扰动检测等。E-mail: yufeng3378@163.com
　　余　艳(1980-),女,通信作者,博士研究生,副教授,研究方向为计算机视觉、机器学习。E-mail:yuyan@wust.edu.cn